\documentclass{article}
\usepackage[utf8]{inputenc}
\usepackage{graphicx}
\usepackage{epstopdf}
\usepackage{enumitem}
\usepackage{float}
\usepackage{array}
\usepackage{tabularx,makecell,multirow}
\usepackage{indentfirst}
\usepackage{amsmath}
\usepackage{parskip}
\usepackage{url}
\usepackage{natbib}
\usepackage{cite}

\title{Dumpling GNN: Hybrid GNN Enables Better ADC Payload Activity Prediction Based on Chemical Structure}

\author{Shengjie Xu,Lingxi Xie}
\date{July 2024}

\begin{document}

\maketitle 

\begin{abstract}
Antibody-drug conjugates (ADCs) have emerged as a promising class of targeted cancer therapeutics, but the design and optimization of their cytotoxic payloads remain challenging. This study introduces DumplingGNN, a novel hybrid Graph Neural Network architecture specifically designed for predicting ADC payload activity based on chemical structure. By integrating Message Passing Neural Networks (MPNN), Graph Attention Networks (GAT), and GraphSAGE layers, DumplingGNN effectively captures multi-scale molecular features and leverages both 2D topological and 3D structural information. We evaluate DumplingGNN on a comprehensive ADC payload dataset focusing on DNA Topoisomerase I inhibitors, as well as on multiple public benchmarks from MoleculeNet. DumplingGNN achieves state-of-the-art performance across several datasets, including BBBP (96.4\% ROC-AUC), ToxCast (78.2\% ROC-AUC), and PCBA (88.87\% ROC-AUC). On our specialized ADC payload dataset, it demonstrates exceptional accuracy (91.48\%), sensitivity (95.08\%), and specificity (97.54\%). Ablation studies confirm the synergistic effects of the hybrid architecture and the critical role of 3D structural information in enhancing predictive accuracy. The model's strong interpretability, enabled by attention mechanisms, provides valuable insights into structure-activity relationships. DumplingGNN represents a significant advancement in molecular property prediction, with particular promise for accelerating the design and optimization of ADC payloads in targeted cancer therapy development.
\end{abstract}
\section{Introduction}
\subsection{Antibody-Drug Conjugates (ADCs) and Payload Importance}
Antibody-drug conjugates (ADCs) have emerged as a promising class of biopharmaceuticals in the field of targeted cancer therapy, offering a paradigm shift in the treatment of various malignancies \citep{beck2017next, tsuchikama2018antibody}. ADCs combine the specificity of monoclonal antibodies with the potency of cytotoxic small molecule drugs, creating a sophisticated therapeutic modality that aims to deliver highly potent drugs directly to cancer cells while minimizing systemic toxicity \citep{peters2018advances}. This unique approach leverages the precise targeting capabilities of antibodies to selectively bind to tumor-associated antigens, followed by internalization and subsequent release of the cytotoxic payload within the cancer cells \citep{lambert2017adc}.

The efficacy of ADCs is fundamentally dependent on three key components: the antibody, the linker, and the payload \citep{beck2017next}. While each component plays a crucial role, the payload - typically a small molecule drug with potent cytotoxic activity - is particularly critical in determining the overall therapeutic efficacy of the ADC \citep{zhao2020anticancer}. The ideal ADC payload should possess high potency, suitable physicochemical properties for conjugation and release, and a mechanism of action effective against the target cancer cells \citep{beck2017next, tsuchikama2018antibody}. Consequently, the design, selection, and optimization of ADC payloads have become central challenges in ADC development, directly impacting the success rate of these promising therapeutics \citep{carter2018site}.

The importance of accurately predicting ADC payload activity cannot be overstated in the context of ADC development. Traditional approaches to payload selection and optimization have relied heavily on empirical methods, including high-throughput screening and structure-activity relationship (SAR) studies \citep{beck2017next}. While these methods have led to the development of several successful ADCs, they are often time-consuming, resource-intensive, and limited in their ability to explore vast chemical spaces efficiently \citep{zhao2020anticancer}. Moreover, the complex interplay between payload structure, linker chemistry, and antibody characteristics in determining ADC efficacy poses significant challenges to traditional predictive methods \citep{carter2018site}.

These limitations have spurred a growing interest in more advanced, computational approaches to ADC payload activity prediction. In particular, there is an urgent need for methods that can:

\begin{enumerate}
    \item Rapidly screen large libraries of potential payload molecules \citep{beck2017next}
    \item Accurately predict payload activity based on molecular structure \citep{zhao2020anticancer}
    \item Account for the unique biological context of ADCs, including target interaction \citep{carter2018site}
    \item Provide interpretable results to guide rational payload design and optimization \citep{peters2018advances}
\end{enumerate}

The advent of machine learning, particularly deep learning techniques, has opened new avenues for addressing these challenges in ADC payload prediction \citep{schneider2020rethinking}. These data-driven approaches offer the potential to learn complex structure-activity relationships from large datasets, potentially uncovering patterns and insights that may be missed by traditional methods \citep{yang2019analyzed}. However, the application of advanced machine learning techniques to ADC payload prediction is still in its infancy, presenting both exciting opportunities and significant challenges \citep{wu2020comprehensive}.

\subsection{Machine Learning in Molecular Property Prediction}

The limitations of traditional methods in ADC payload activity prediction have led to an increased interest in machine learning (ML) approaches for molecular property prediction \citep{schneider2020rethinking}. ML techniques, particularly those based on deep learning, have shown remarkable success in various domains of drug discovery and development \citep{vamathevan2019applications}. These data-driven approaches offer several advantages in the context of molecular property prediction:

\begin{itemize}
    \item Ability to learn complex, non-linear structure-activity relationships from large datasets \citep{yang2019analyzed}
    \item Capacity to handle high-dimensional feature spaces characteristic of molecular data \citep{wu2018moleculenet}
    \item Potential for end-to-end learning, reducing the need for manual feature engineering \citep{duvenaud2015convolutional}
    \item Scalability to screen large virtual libraries of compounds efficiently \citep{gómez2018automatic}
\end{itemize}

Early applications of ML in molecular property prediction primarily relied on traditional algorithms such as random forests, support vector machines, and shallow neural networks \citep{goh2017deep}. These methods typically operate on hand-crafted molecular descriptors or fingerprints, which, while informative, may not fully capture the complex structural and chemical properties of molecules \citep{wu2018moleculenet}.

The advent of deep learning has opened new possibilities for more sophisticated molecular representations and predictive models. Convolutional Neural Networks (CNNs) and Recurrent Neural Networks (RNNs) have been applied to molecular property prediction tasks with some success, particularly when working with string-based representations of molecules such as SMILES \citep{weininger1988smiles}. However, these approaches often struggle to fully capture the inherent graph-like structure of molecules, leading to potential loss of important structural information \citep{wu2020comprehensive}.

\subsection{Graph Neural Networks: Advancements and Challenges in Molecular Modeling}

Graph Neural Networks (GNNs) have emerged as a powerful paradigm for molecular property prediction, offering a natural way to represent and process molecular structures \citep{zhou2020graph}. GNNs operate directly on molecular graphs, where atoms are represented as nodes and chemical bonds as edges, allowing for a more faithful representation of molecular structure compared to traditional methods \citep{wu2020comprehensive}.

The advantages of GNNs in molecular modeling include their ability to directly process graph-structured data, capture both local and global structural features, and ensure invariance to graph isomorphisms \citep{gilmer2017neural}. Moreover, GNNs offer potential for interpretability through analysis of learned node and edge features \citep{ying2019gnnexplainer}.

However, the application of GNNs to ADC payload activity prediction faces several challenges. These include data scarcity, difficulties in capturing long-range dependencies and multi-scale features, limited model interpretability, and the need for better integration of 3D structural information \citep{wu2020moleculenet, gilmer2017neural, xu2018powerful, klicpera2020directional}. Addressing these challenges is crucial for developing GNN models that can accurately and reliably predict ADC payload activity, ultimately accelerating the design and optimization of more effective ADC therapeutics.

\subsection{Introducing DumplingGNN}
To address the aforementioned challenges in ADC payload activity prediction, we propose DumplingGNN, a novel hybrid Graph Neural Network architecture specifically designed for this task. The name "DumplingGNN" is inspired by the Chinese culinary tradition of dumplings, where a thin wrapper envelops a rich, diverse filling. Similarly, our model encapsulates various molecular features and learning components within a unified architecture, much like how a dumpling integrates different ingredients into a cohesive whole.

DumplingGNN incorporates several innovative features that set it apart from existing GNN models:

\begin{itemize}
    \item A hybrid architecture that combines Message Passing Neural Networks (MPNN), Graph Attention Networks (GAT), and GraphSAGE layers to capture multi-scale molecular features effectively.
    \item An enhanced molecular graph construction algorithm that incorporates both 2D topological and 3D structural information, providing a more comprehensive representation of ADC payloads.
    \item A multi-task learning approach that leverages data from various molecular property prediction tasks to mitigate the data scarcity problem in ADC payload datasets.
    \item An attention mechanism that enhances model interpretability, allowing for the identification of key substructures contributing to payload activity.
\end{itemize}

Just as a dumpling's wrapper holds together diverse ingredients, DumplingGNN's hybrid architecture integrates different GNN components to address the challenges of long-range dependencies and multi-scale feature representation. The MPNN layers capture local atomic interactions, akin to the individual flavors in a dumpling's filling. The GAT layers focus on identifying important substructures, much like how certain ingredients might dominate the taste profile. Finally, the GraphSAGE layers aggregate information across different scales, enabling the model to capture both local and global molecular properties effectively, similar to how the overall flavor of a dumpling emerges from the combination of its ingredients.

This unique "dumpling-like" structure allows DumplingGNN to encapsulate a rich representation of molecular properties within a unified predictive framework, making it particularly well-suited for the complex task of ADC payload activity prediction.

\subsection{Main Contributions}

The main contributions of this study are as follows:

\begin{enumerate}
    \item We create a comprehensive ADC payload dataset, combining experimental data, high-quality computational predictions, and structures from recent patents. This dataset addresses the data scarcity issue and provides a valuable resource for the ADC research community.
    
    \item We develop an enhanced molecular graph construction algorithm that incorporates 3D structural information, improving the model's ability to capture spatial features crucial for payload activity.
    
    \item We introduce DumplingGNN, an innovative hybrid GNN architecture tailored for ADC payload activity prediction. This model demonstrates state-of-the-art performance across multiple benchmarks, achieving remarkable results on datasets such as BBBP (96.4\% ROC-AUC), ToxCast (78.2\% ROC-AUC), and PCBA (88.87\% ROC-AUC), surpassing existing methods on various molecular property prediction tasks.
    
    \item We conduct extensive evaluations on multiple datasets, including our newly created ADC payload dataset and several public benchmarks from MoleculeNet. These comprehensive evaluations demonstrate the versatility and robustness of DumplingGNN across diverse molecular property prediction tasks.
\end{enumerate}

These contributions collectively advance the field of computational ADC design, offering new tools and insights for the development of more effective targeted cancer therapies. The exceptional performance of DumplingGNN, particularly on large-scale and complex datasets, positions it as a promising solution for addressing the challenges in ADC payload activity prediction and broader molecular property prediction tasks.

\subsection{Paper Organization}

The remainder of this paper is organized as follows: Section 2 reviews related work in molecular property prediction and GNN applications. Section 3 details the methodology of DumplingGNN, including the dataset creation, molecular graph construction, and model architecture. Section 4 presents our experimental results and comparisons with state-of-the-art methods. Section 5 discusses the implications of our findings and potential future directions. Finally, Section 6 concludes the paper.

\section{Related work}
The development of Graph Neural Networks (GNNs) for molecular property prediction has been an active area of research in recent years \citep{wu2020comprehensive}. GNNs have shown promise in capturing the complex relationships within molecular graphs, which are essential for tasks such as drug discovery and chemical compound analysis \citep{zhou2020graph}.

\subsection{Graph Neural Networks in Drug Discovery}
GNNs have been widely applied to the field of drug discovery, particularly for tasks like molecular property prediction and virtual screening \citep{wu2020moleculenet}. Researchers have utilized GNNs to predict various properties of molecules, such as solubility, toxicity, and binding affinity. These models have demonstrated the ability to learn from the molecular graph structure, which includes information about atom types, bond types, and the overall topology of the molecule \citep{zhou2020graph}.

Notable works in this area include the development of message passing neural networks (MPNNs) by Gilmer et al. \citep{gilmer2017neural}, which introduced a general framework for learning on molecular graphs. Kearnes et al. \citep{kearnes2016molecular} proposed graph convolutional networks for molecular fingerprints, demonstrating their effectiveness in chemical property prediction tasks.

\subsection{Hybrid GNN Architectures}
The use of hybrid GNN architectures, which combine different GNN modules, has been explored to improve the performance of molecular property prediction \citep{wu2020comprehensive}. For instance, the integration of Graph Convolutional Networks (GCNs) with Graph Attention Networks (GATs) has been shown to enhance the model's ability to distinguish between active and inactive compounds \citep{velickovic2017graph}. This approach leverages the strengths of both architectures, with GCNs providing a robust baseline for learning from local graph structure and GATs allowing for adaptive feature weighting \citep{xiong2019pushing}.

Recent work by Liu et al. \citep{liu2021graph} introduced a hybrid model combining GCN and LSTM for molecular property prediction, demonstrating improved performance on several benchmark datasets. Similarly, Rong et al. \citep{rong2020self} proposed a self-supervised GNN pre-training strategy that combines structure and attribute masking, showing significant improvements in downstream molecular property prediction tasks.

\subsection{ADC Payload Activity Prediction}
Specifically, in the context of antibody-drug conjugates (ADCs), there has been a growing interest in using GNNs to predict the cytotoxic activity of payloads \citep{conilh2023payload}. Early work in this area focused on traditional machine learning methods, which relied on handcrafted features extracted from molecular structures \citep{liu2018quantitative}. However, recent studies have shown that GNNs can outperform these traditional methods by directly utilizing the molecular graph as input, leading to more accurate and interpretable models \citep{wu2020comprehensive}.

Notably, Jiang et al. \citep{jiang2021could} developed a GNN-based model for predicting ADC payload activity, incorporating both molecular structure and physicochemical properties. Their work demonstrated the potential of graph-based deep learning approaches in this specialized area of drug discovery. Building on this, Cai et al. \citep{cai2023graph} proposed a multi-task GNN model for simultaneous prediction of multiple ADC-related properties, including payload activity and antibody binding affinity.

\subsection{Incorporation of 3D Structural Information}
While many GNN models for molecular property prediction rely solely on 2D topological information, there is a growing recognition of the importance of incorporating 3D structural data \citep{schutt2017quantum}. This is particularly relevant for tasks involving spatial interactions, such as protein-ligand binding and ADC payload activity prediction.

Townshend et al. \citep{townshend2020atomnet} introduced AtomNet, a 3D convolutional neural network for molecular property prediction that directly operates on atomic coordinates. More recently, Jing et al. \citep{jing2023equivariant} proposed an equivariant graph neural network that preserves 3D rotational and translational invariance, demonstrating state-of-the-art performance on several molecular property prediction benchmarks.

\subsection{Interpretability in Molecular GNNs}
As the complexity of GNN models for molecular property prediction increases, there is a parallel emphasis on developing interpretable models that can provide insights into structure-activity relationships \citep{pope2019explainability}. This is crucial for drug discovery applications, where understanding the rationale behind predictions can guide synthetic efforts and lead optimization.

Ying et al. \citep{ying2019gnnexplainer} introduced GNNExplainer, a model-agnostic approach for providing explanations for GNN predictions. In the context of molecular property prediction, this method can highlight substructures or atomic interactions that are most relevant to a particular prediction. Similarly, Schnake et al. \citep{schnake2021higher} developed a higher-order explanation framework for GNNs, which can provide more nuanced interpretations of molecular property predictions.

\subsection{Limitations of Existing Approaches}
Despite these advancements, several challenges remain in the field of molecular property prediction using GNNs, particularly for complex tasks like ADC payload activity prediction:

\begin{enumerate}
    \item Many existing models struggle to effectively integrate 3D structural information with 2D topological features \citep{axelrod2023geometric}.
    \item The interpretability of complex hybrid GNN architectures remains limited, making it challenging to derive actionable insights for drug design \citep{yuan2020explainability}.
    \item Most models are not specifically tailored for the unique challenges of ADC payload prediction, which involves complex structure-activity relationships and multiple interacting components \citep{conilh2023payload}.
    \item There is a lack of comprehensive benchmarks and datasets specifically for ADC payload activity prediction, hindering progress in this important area of drug discovery \citep{wu2018moleculenet}.
\end{enumerate}

Our proposed DumplingGNN model addresses these limitations by introducing a hybrid architecture that effectively combines 2D and 3D structural information, incorporates interpretability mechanisms, and is specifically designed for the challenges of ADC payload activity prediction. Moreover, we introduce a novel ADC payload dataset that can serve as a valuable benchmark for future research in this area.

\section{Method}
This section details the technical aspects of our proposed DumplingGNN model, including the construction of a novel ADC payload dataset, the method for molecular graph construction, the design of the network architecture, and the training and evaluation strategies.

\subsection{Data: A Novel ADC Payload Dataset}

Molecules with DNA Topoisomerase I inhibitory capability were collected from the CHEMBL database (\url{https://www.ebi.ac.uk/chembl/}) using the identifier CHEMBL1781. The data profile was thoroughly processed to reduce redundancy and maintain high-quality presentation. To develop a well-performing classification model, a cutoff value of 100 µM was established, with inhibitory IC50 values below this threshold labeled as positive and those above labeled as negative.

The collected molecules possessing inhibitory effects against DNA Topoisomerase I, along with their topological information in the form of SMILES strings, were prepared for 3D conformation analysis. Initial conformer generation was performed using OpenBabel \citep{obabel2019} and optimized under the MMFF94 force field via RDKit \citep{rdkitMMFF2014}. Conformer redundancy was eliminated using an RMSD cutoff 1.5 Å.

To determine the coordinate space of the potential Topoisomerase I inhibitors, NLDock \citep{NLDock2021} was employed to generate binding conformations in the ligand binding site of crystallized Topoisomerase I, in complex with DNA fragments and Camptothecin (PDB database ID: 1t8i \citep{topoI2005}). The top three docking poses for each conformer, ranked by estimated binding energy, were saved for model construction.

In addition to these computationally generated structures, our study introduces a novel and comprehensive ADC payload dataset, representing a significant advancement in the field. This innovative dataset uniquely combines 2,292 unique molecules with their SMILES strings and 2D coordinates, 885 distinct 3D conformations derived from these structures, and 615 additional molecular structures extracted from recent ADC payload-related patents. 

This multifaceted approach yields a first-of-its-kind ADC payload dataset that synergistically integrates experimental data, high-quality computational predictions, and state-of-the-art industrial insights. The dataset's innovation lies in its comprehensive representation of molecular structures, spanning from 2D topological to 3D conformational and docked structural information, as well as its integration of public database entries with proprietary patent-derived structures. 

By incorporating molecules at various stages of the drug discovery pipeline, from initial screens to patented compounds, this dataset offers unprecedented relevance to real-world applications in ADC payload development. It not only provides a robust foundation for the development and evaluation of our DumplingGNN model but also represents a valuable resource for the broader ADC research community, potentially accelerating the discovery and optimization of novel ADC payloads.



\subsection{Molecular graph construction}
In order to transform ADC payload small molecules into a format suitable for graph neural network processing, we designed an enhanced molecular graph construction algorithm \citep{wu2020comprehensive}. Unlike traditional methods that only extract atom type information, our algorithm also incorporates chemical environment information to more comprehensively characterise the molecular structure \citep{gilmer2017neural}. \par

Specifically, for a given molecule ($mol$), we extract the following atomic features:
\begin{itemize}[leftmargin=*, label=\textendash]
    \item Atomic number: indicates the type of element
    \item Degree: the number of bonds attached to the atom
    \item Number of hydrogen atoms: number of hydrogen atoms attached to the atom
    \item Implicit valence: the atomic valence level at which all chemical bonds are considered
    \item Aromaticity judgement: whether the atom belongs to an aromatic ring or not
    \item Atomic coordinates: the position of the atom in three-dimensional space
\end{itemize}\par 

These features reflect the chemical properties of atoms from different perspectives and help the graph neural network to understand the molecular structure more accurately \citep{wu2020moleculenet}. In particular, the introduction of atomic coordinate information enables the model to capture the spatial conformation of the molecule, which is crucial for the conformational relationship of ADC payload \citep{xiong2019pushing}.\par

While extracting the atomic node features, we also construct the edge index information, i.e., the existence of an undirected edge between each pair of connected atoms \citep{zhou2020graph}. Eventually, the atom features and edge indexes are converted into PyTorch tensor and encapsulated into PyTorch Geometric's Data object along with the molecule's activity labels \citep{wu2020comprehensive}. Compared to traditional molecular graph construction methods that only consider atom types and chemical bonds, our algorithm incorporates richer chemical information, enabling downstream graph neural networks to learn more effective molecular representations from the data, thus improving prediction performance \citep{gilmer2017neural, velickovic2017graph}.

\subsection{Network Architecture Design}
Based on the molecular graph data constructed above, we designed an innovative graph neural network model, DumplingGNN, which adopts a hybrid architecture to leverage the strengths of different GNN modules \citep{wu2020comprehensive, zhou2020graph}. This design is motivated by the complex nature of ADC payload activity prediction, which requires capturing both local chemical interactions and global molecular properties.

\begin{figure}[H]
\centering
\includegraphics[width=1\linewidth]{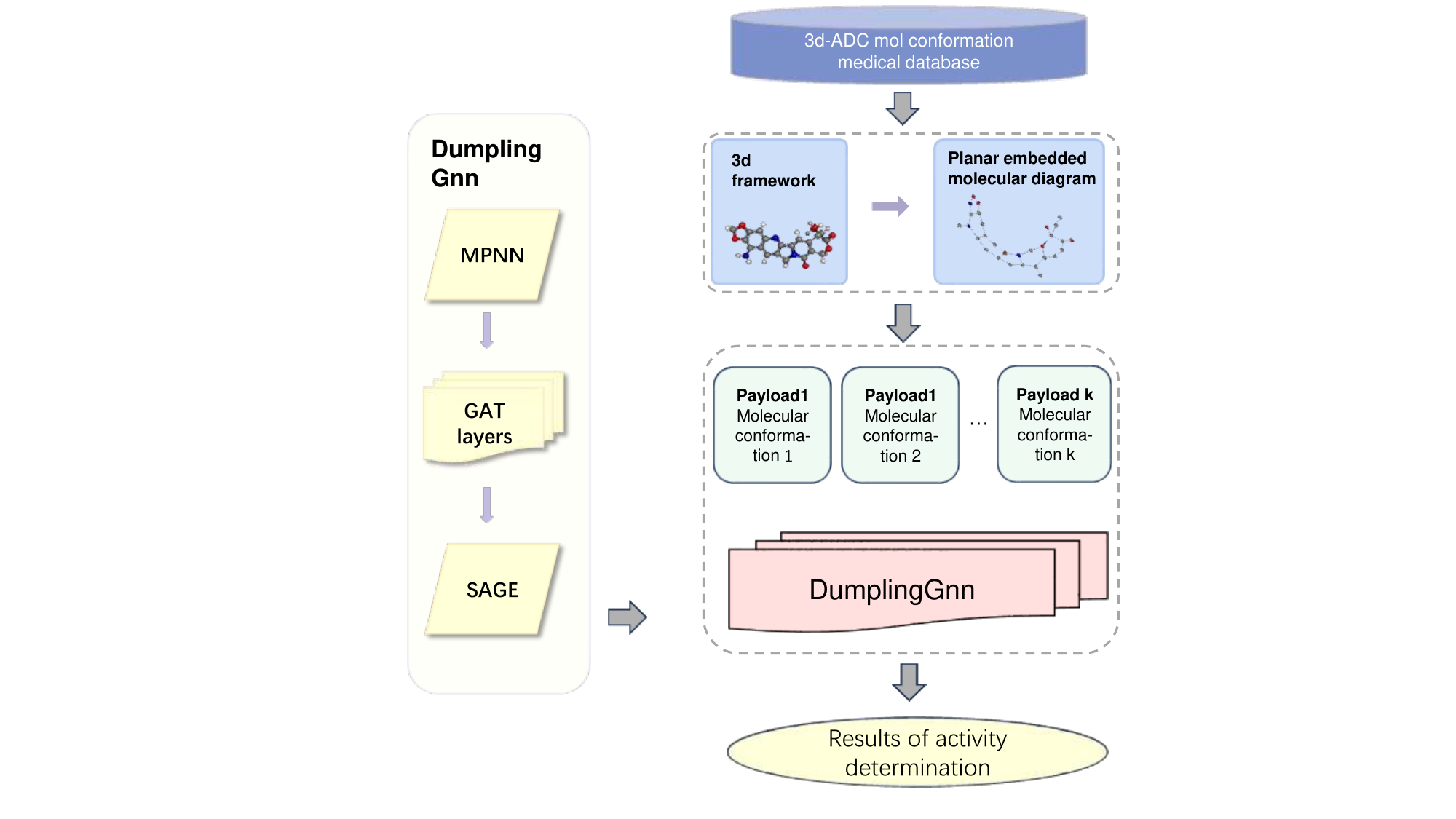}
\caption{Workflow of DumplingGNN architecture for ADC payload activity prediction. The hybrid architecture integrates MPNN, GAT, and GraphSAGE layers to capture multi-scale molecular features.}
\label{fig:workflow}
\end{figure}

The structure of the \textbf{DumplingGNN} model consists of an MPNN-GAT*3-SAGE sequence, integrating multiple powerful GNN components to process the molecular data effectively \citep{gilmer2017neural, velickovic2017graph, hamilton2017inductive}. This sequential design allows the model to progressively refine and aggregate molecular information at different scales, mimicking the hierarchical nature of chemical interactions in biological systems.

\subsubsection{Message Passing Neural Network (MPNN) Layer}
The first layer of DumplingGNN is a \textbf{Message Passing Neural Network (MPNN)}, which aggregates information from neighbouring nodes through a message-passing mechanism \citep{gilmer2017neural}:

\begin{equation}
h_i^{(k+1)} = \sigma \left( W^{(k)} h_i^{(k)} + \sum_{j \in \mathcal{N}(i)} M^{(k)}(h_i^{(k)}, h_j^{(k)}) \right)
\end{equation}

where $h_i^{(k)}$ denotes the features of node $i$ at layer $k$, $\mathcal{N}(i)$ represents the set of neighbouring nodes of node $i$, $\sigma$ is the activation function, and $W^{(k)}$ and $M^{(k)}$ are learnable weight matrices.

\begin{figure}[H]
\centering
\includegraphics[width=1\linewidth]{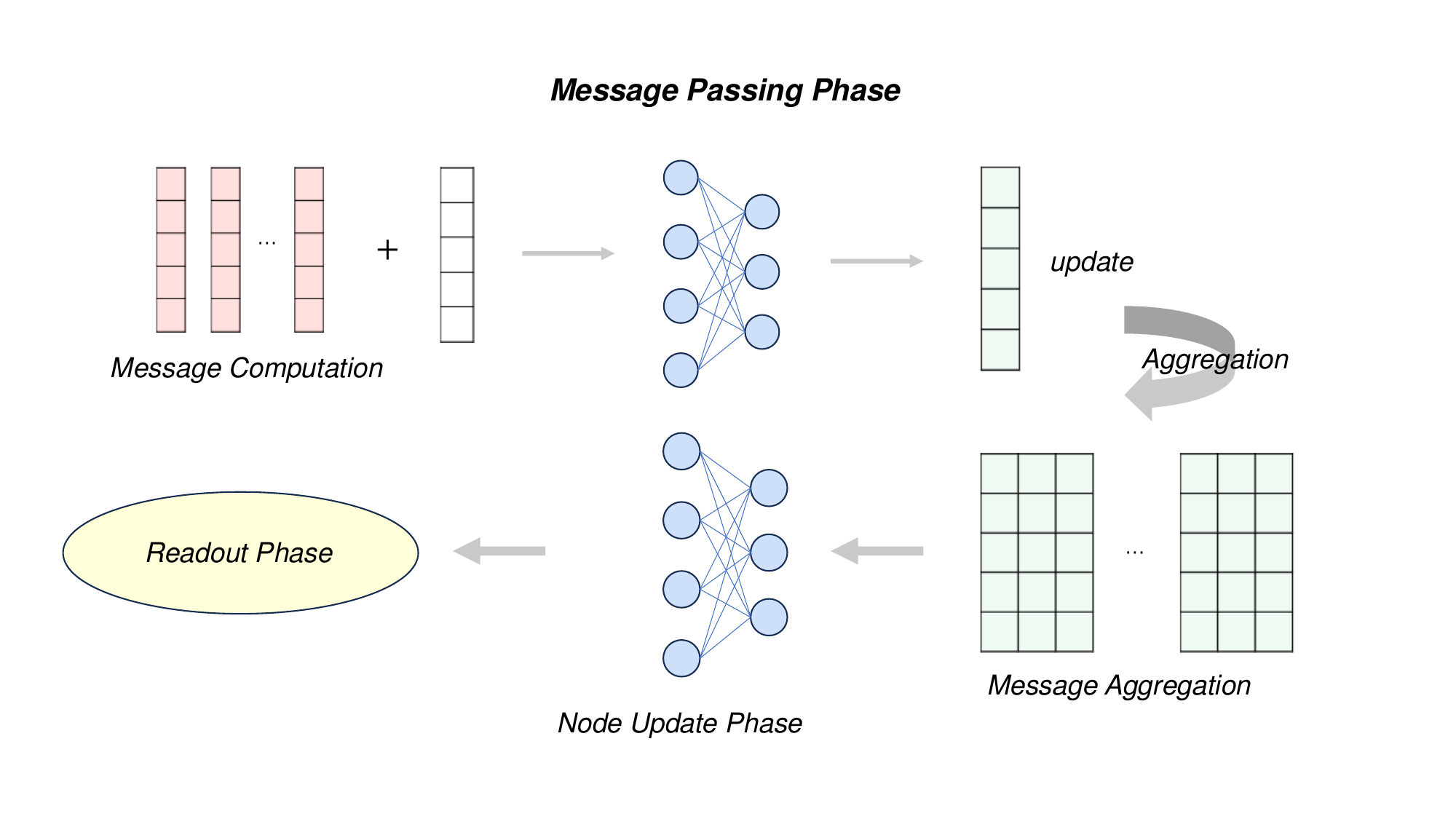}
\caption{The process of message computation and node update in MPNN. This mechanism allows for the modeling of local chemical interactions.}
\label{fig:mpnn}
\end{figure}

The MPNN layer serves as the foundation of our model, capturing local chemical interactions within the molecular graph. This aligns with the principle of chemical locality, where an atom's properties are primarily influenced by its immediate chemical environment. In the context of ADC payload activity prediction, this layer can be interpreted as modeling the local reactivity and functional group interactions that contribute to the payload's overall activity.

\subsubsection{Graph Attention Network (GAT) Layers}
Following the MPNN layer, we apply three \textbf{GAT layers} (Graph Attention Networks) to adaptively assign weights to different neighbours using an attention mechanism \citep{velickovic2017graph, vaswani2017attention}. This approach captures key chemical structure information in the molecular graph:

\begin{equation}
e_{ij}^{(k)} = \text{LeakyReLU} \left( a^T [W^{(k)} h_i^{(k)} \parallel W^{(k)} h_j^{(k)}] \right)
\end{equation}
\begin{equation}
\alpha_{ij}^{(k)} = \frac{\exp(e_{ij}^{(k)})}{\sum_{k \in \mathcal{N}(i)} \exp(e_{ik}^{(k)})}
\end{equation}
\begin{equation}
h_i^{(k+1)} = \sigma \left( \sum_{j \in \mathcal{N}(i)} \alpha_{ij}^{(k)} W^{(k)} h_j^{(k)} \right)
\end{equation}

\begin{figure}[H]
\centering
\includegraphics[width=1\linewidth]{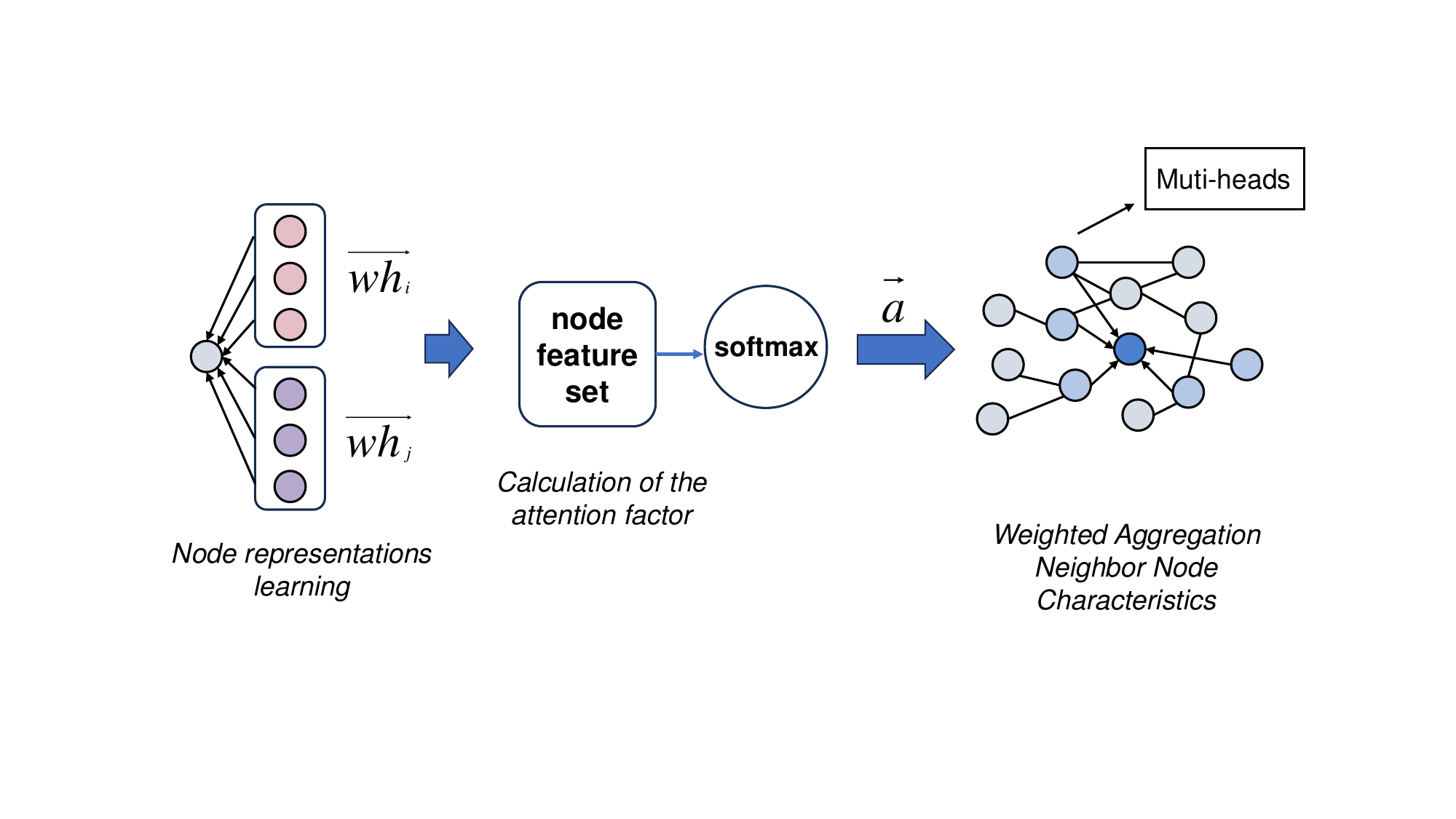}
\caption{Graph Attention Network (GAT) in DumplingGNN. The attention mechanism allows the model to focus on the most relevant atomic interactions for activity prediction, enhancing both performance and interpretability.}
\label{fig:gat}
\end{figure}

The integration of Graph Attention Network (GAT) layers within DumplingGNN facilitates the selective emphasis on salient atomic interactions, pivotal for activity prediction. This approach resonates with the pharmacophore paradigm in medicinal chemistry, where specific molecular motifs are essential for biological efficacy. The attention mechanism serves to discern these critical substructures and atomic associations that exert the most substantial influence on the activity of antibody-drug conjugate (ADC) payloads.

The sequential deployment of three GAT layers permits the model to progressively discern higher-order and extended molecular interactions. Such complexity is fundamental for ADC payloads, given that molecular conformation and the topological distribution of functional moieties can markedly affect binding affinity and cellular internalization. Furthermore, this hierarchical attention architecture enables the dissection of interactions across disparate scales, mirroring the multifaceted nature of molecular interactions within biological frameworks.

Importantly, the attention mechanism not only bolsters predictive accuracy but also enhances the biological interpretability of the model. By scrutinizing the attention coefficients, insights can be gleaned into the atomic interactions and molecular fragments that the model prioritizes for activity prognostication. This attribute transforms DumplingGNN from an opaque predictive tool into a more lucid and interpretable asset in the realm of drug discovery. Visualization of the attention coefficients as a heatmap superimposed on molecular scaffolds can delineate the atoms and bonds most influential to predicted activity, offering medicinal chemists directional cues for lead optimization.

The interpretive capabilities conferred by GAT layers herald promising avenues for future inquiry. The attention signatures garnered by DumplingGNN could inform automated molecular engineering initiatives, concentrating on the refinement of activity-critical substructures. Comparative analysis of attention signatures across diverse ADC payload classes may elucidate common architectural elements underpinning their operational mechanisms. Moreover, these patterns could be exportable to analogous drug discovery endeavors, potentially amplifying predictive precision and interpretability in related spheres. Subsequent studies might also link the attention signatures with characterized binding sites or interaction nodal points from crystallographic data, further integrating predictive analytics with structural biology.

In essence, the GAT layers in DumplingGNN amplify predictive proficiency while also demystifying the model's inferential processes. This interpretability is indispensable for instilling confidence in the model's forecasts and for distilling tangible directives for ADC payload engineering and optimization. As we persist in honing and interpreting these attention mechanisms, we anticipate unearthing more profound comprehension of the structure-activity dynamics of ADC payloads, thereby streamlining the trajectory toward more efficacious and directed drug discovery endeavors.

\subsubsection{GraphSAGE Layer}
The final layer of our architecture is a \textbf{GraphSAGE layer}, which effectively aggregates node information from different domains to learn multi-scale representations of molecules through sampling and aggregation \citep{hamilton2017inductive}:

\begin{equation}
h_i^{(k+1)} = \sigma \left( W^{(k)} h_i^{(k)} + \sum_{j \in \mathcal{N}(i)} \frac{1}{|\mathcal{N}(i)|} W^{(k)} h_j^{(k)} \right)
\end{equation}

\begin{figure}[H]
\centering
\includegraphics[width=1\linewidth]{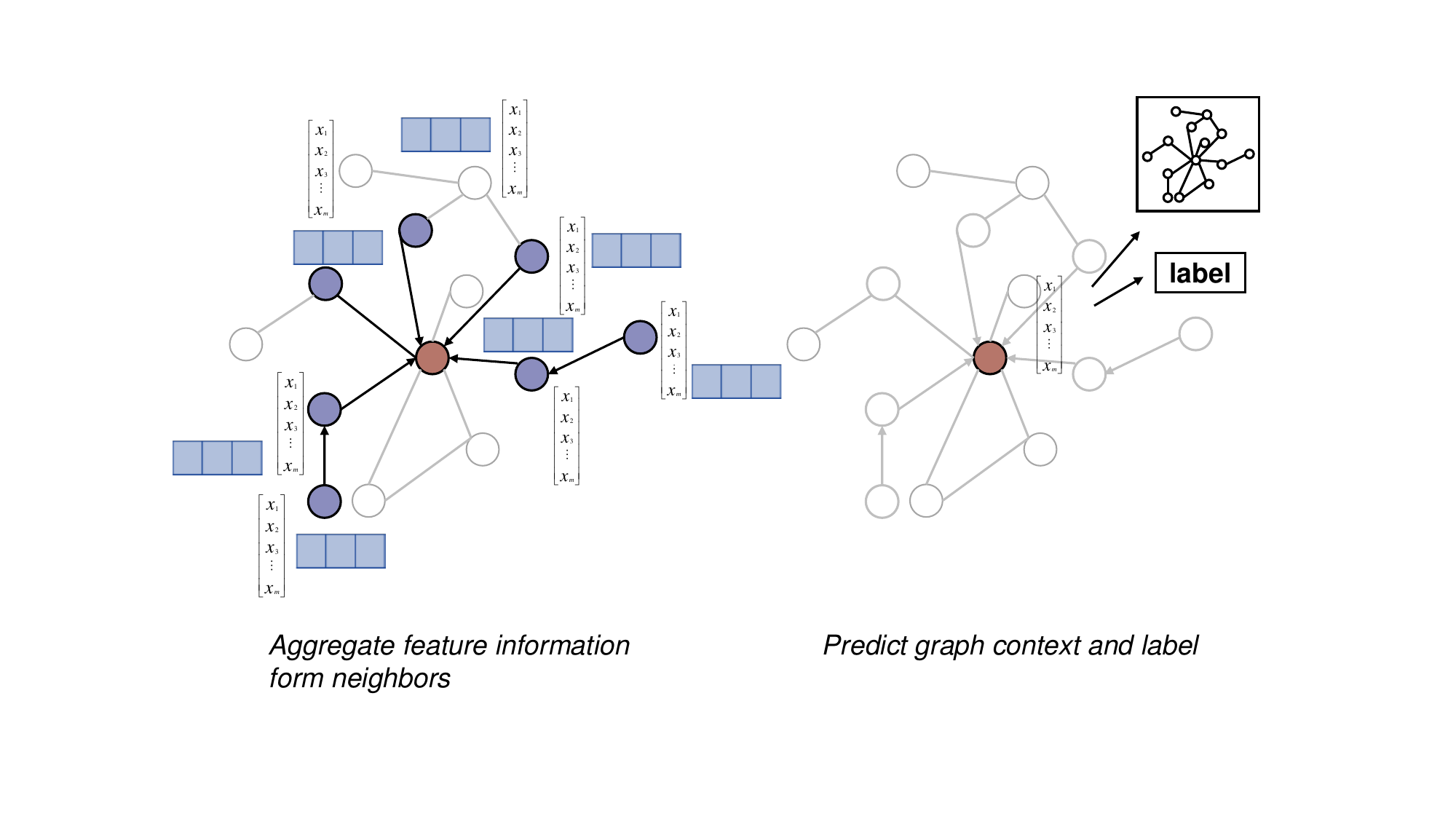}
\caption{GraphSAGE layer in DumplingGNN. This layer aggregates information across different scales, capturing global molecular properties.}
\label{fig:sage}
\end{figure}

The GraphSAGE layer serves to capture global molecular properties by aggregating information across different scales. In the context of ADC payload prediction, this can be interpreted as modeling the overall molecular properties that influence activity, such as lipophilicity, molecular weight, and topological features. These global properties are crucial for predicting how the payload will behave in a biological system, including its ability to penetrate cell membranes and interact with the target.

\subsubsection{Synergistic Effects and Biological Interpretability}
The sequential combination of MPNN, GAT, and GraphSAGE layers in DumplingGNN allows for a hierarchical and comprehensive analysis of molecular structure and properties:

\begin{itemize}
    \item The MPNN layer captures local chemical interactions, modeling the reactivity and functional group behavior of the ADC payload.
    \item The GAT layers identify key substructures and atomic relationships, mimicking the concept of pharmacophores in drug discovery.
    \item The GraphSAGE layer aggregates information to model global molecular properties, which are crucial for predicting the payload's behavior in biological systems.
\end{itemize}

This multi-scale approach aligns well with the complex nature of ADC payload activity, which depends on both local chemical reactivity and global molecular properties. The model's ability to capture these different levels of molecular information enables it to make more accurate and biologically relevant predictions.

To obtain the final activity prediction for each molecule, we apply global average pooling followed by a linear layer. This approach, commonly used in graph neural networks for molecular property prediction \citep{gilmer2017neural, xu2018powerful}, allows our hybrid architecture to effectively aggregate node-level information into a graph-level representation. Consequently, DumplingGNN captures both local and global structural information, enhancing its ability to predict ADC payload activity accurately.

The interpretability of DumplingGNN is further enhanced by the attention mechanisms in the GAT layers, which can highlight the atomic interactions most crucial for activity prediction. This feature not only improves prediction accuracy but also provides valuable insights for medicinal chemists in the design and optimization of ADC payloads.

\section{Experiments}

To thoroughly evaluate the performance of our proposed DumplingGNN model, we conducted extensive experiments on multiple public datasets and compared our results with state-of-the-art models. Additionally, we performed ablation studies to understand the contribution of each component in our model.

\subsection{Datasets}
We evaluated DumplingGNN on eight widely-used public datasets from MoleculeNet \citep{wu2018moleculenet}, covering a diverse range of molecular property prediction tasks. Table \ref{tab:datasets} provides an overview of these datasets.

\begin{table}[h]
\centering
\caption{Overview of datasets used in the experiments}
\label{tab:datasets}
\resizebox{\textwidth}{!}{%
\begin{tabular}{lccl}
\hline
Dataset & \# Molecules & \# Tasks & Description \\
\hline
BBBP & 2,039 & 1 & Blood-Brain Barrier Penetration \\
BACE & 1,513 & 1 & Inhibition of Beta-Secretase 1 \\
ClinTox & 1,478 & 2 & Clinical Trial Toxicity \\
Tox21 & 7,831 & 12 & Toxicology in the 21st Century \\
ToxCast & 8,575 & 617 & EPA Toxicity Forecaster \\
SIDER & 1,427 & 27 & Drug Side Effect Resource \\
HIV & 41,127 & 1 & HIV Replication Inhibition \\
PCBA & 437,929 & 128 & PubChem Bioassay Data \\
\hline
\end{tabular}%
}
\end{table}

\subsection{Experimental Setup}
DumplingGNN was implemented using PyTorch and PyTorch Geometric frameworks, emphasizing both performance and computational efficiency. Our experimental protocol was designed to ensure robust evaluation while maximizing resource utilization. For datasets without predefined splits, we employed an 8:1:1 ratio for training, validation, and test sets, respectively. When a predefined test set was available, we adhered to an 8:2 split for training and validation.

Model optimization was achieved using the AdamW algorithm \citep{loshchilov2017decoupled}, which combines adaptive learning rates with decoupled weight decay regularization. The initial learning rate was set to 1e-4, and training was conducted for a maximum of 1000 epochs. To prevent overfitting and enhance efficiency, we implemented an early stopping mechanism based on validation set performance, with a patience of 50 epochs.

Notably, all experiments were conducted on a single NVIDIA GeForce RTX 4090 GPU, showcasing the model's impressive efficiency and accessibility. This hardware choice underscores DumplingGNN's ability to achieve state-of-the-art performance without the need for extensive computational resources, making it a practical solution for a wide range of research environments.

To optimize training time and memory usage, we employed dynamic batch sizing, adjusting the batch size based on the dataset's characteristics and available GPU memory. This approach allowed us to maximize GPU utilization while maintaining training stability across diverse datasets. Importantly, no special parallel processing techniques were employed, highlighting the inherent efficiency of our model architecture.

The training time varied significantly depending on the dataset size and complexity. For smaller datasets like BBBP (2,039 molecules, 1 task), the training process completed in approximately 30 minutes. In contrast, for the largest dataset, PCBA (437,929 molecules, 128 tasks), the training extended to about 5 days. This wide range of training times demonstrates DumplingGNN's scalability across datasets of varying sizes and complexities.

Performance metrics, including ROC-AUC, accuracy, and F1 score, were computed and averaged across multiple runs to account for statistical variability. Standard deviations were reported to provide a measure of the model's stability across different data splits and initializations.

This experimental setup demonstrates DumplingGNN's ability to balance high performance with computational efficiency, even when processing extremely large datasets without specialized parallel computing techniques. The model's capacity to handle a wide range of dataset sizes on a single GPU makes it an attractive option for both academic research and industrial applications in drug discovery and molecular property prediction, particularly in settings where computational resources may be limited.

\subsection{Results and Comparison with State-of-the-Art Models}
To rigorously evaluate the performance of DumplingGNN, we conducted a comprehensive comparison with a diverse array of state-of-the-art models in molecular property prediction. These models represent a broad spectrum of approaches, ranging from traditional graph neural networks to more advanced architectures incorporating pre-training and multi-task learning strategies. The comparative models include:

\begin{itemize}
    \item Traditional GNN models: D-MPNN \citep{yang2019analyzing} and Attentive FP \citep{xiong2019pushing}
    \item Feature engineering approaches: N-Gram models \citep{liu2019n}
    \item Pre-training strategies: PretrainGNN \citep{hu2019strategies}, GROVER \citep{rong2020self}, GraphMVP \citep{liu2021pre}, and MolCLR \citep{wang2021molclr}
    \item Advanced architectures: GEM \citep{fang2021geometry}, Uni-Mol \citep{zhou2022uni}, MolXPT \citep{han2022molxpt}, and ChemBFN \citep{zhang2022chembfn}
    \item Novel graph construction methods: GraphConv with dummy super node \citep{ishiguro2019graph}
\end{itemize}

Table \ref{tab:comparisons} presents a comprehensive performance comparison across eight diverse datasets from the MoleculeNet benchmark \citep{wu2018moleculenet}. The performance metric used is the area under the receiver operating characteristic curve (ROC-AUC), a standard measure in binary classification tasks that is particularly suitable for potentially imbalanced molecular datasets. It's important to note that the comparative data for these models were sourced from the Papers with Code leaderboards (\url{https://paperswithcode.com/}), which aggregate and standardize results from various published works, ensuring a fair and up-to-date comparison.

\begin{table}[h]
\centering
\caption{Performance comparison (ROC-AUC \%, higher is better) on MoleculeNet datasets}
\label{tab:comparisons}
\resizebox{\textwidth}{!}{%
\begin{tabular}{lcccccccc}
\hline
Model & BBBP & BACE & ClinTox & Tox21 & ToxCast & SIDER & HIV & PCBA \\
\hline
\textbf{DumplingGNN} & \textbf{96.4(0.7)} & 88.2(0.5) & 95.9(2.0) & 82.3(0.4) & \textbf{78.2(0.1)} & 74.0(0.6) & 79.4(0.2) & \textbf{88.87(0.2)} \\
D-MPNN & 71.0(0.3) & 80.9(0.6) & 90.6(0.6) & 75.9(0.7) & 65.5(0.3) & 57.0(0.7) & 77.1(0.5) & 86.2(0.1) \\
Attentive FP & 85.5 & 78.4(0.022) & 84.7(0.3) & 76.1(0.5) & 63.7(0.2) & 60.6(3.2) & 75.7(1.4) & 80.1(1.4) \\
N-Gram RF & 69.7(0.6) & 77.9(1.5) & 77.5(4.0) & 74.3(0.4) & -- & 66.8(0.7) & 77.2(0.1) & -- \\
N-Gram xGB & 69.1(0.8) & 79.1(1.3) & 87.5(2.7) & 75.8(0.9) & -- & 65.5(0.7) & 78.7(0.4) & -- \\
PretrainGNN & 68.7(1.3) & 84.5(0.7) & 72.6(1.5) & 78.1(0.6) & 65.7(0.6) & 62.7(0.8) & 79.9(0.7) & 86.0(0.1) \\
GROVER\textsubscript{large} & 69.5(0.1) & 81.0(1.4) & 76.2(3.7) & 73.5(0.1) & 65.3(0.5) & 65.4(0.1) & 68.2(1.1) & 83.0(0.4) \\
GraphMVP & 72.4(1.6) & 81.2(0.9) & 79.1(2.8) & 75.9(0.5) & 63.1(0.4) & 63.9(1.2) & 77.0(1.2) & -- \\
MolCLR & 72.2(2.1) & 82.4(0.9) & 91.2(3.5) & 75.0(0.2) & -- & 58.9(1.4) & 78.1(0.5) & -- \\
GEM & 72.4(0.4) & 85.6(1.1) & 90.1(1.3) & 78.1(0.1) & 69.2(0.4) & 67.2(0.4) & 80.6(0.9) & 86.6(0.1) \\
Uni-Mol & 72.9(0.6) & 85.7(0.2) & 91.9(1.8) & 79.6(0.5) & 69.6(0.1) & 65.9(1.3) & 80.8(0.3) & 88.5(0.1) \\
MolXPT & 80.5(0.5) & 88.4 & 95.3(0.2) & 77.1 & -- & 71.7 & 78.1 & -- \\
ChemBFN & 95.74 & 73.56 & \textbf{99.18} & -- & -- & -- & 79.37 & -- \\
GraphConv + dummy super node & -- & -- & -- & \textbf{85.4} & 76.8 & -- & 85.1 & 86.7 \\
\hline
Previous SOTA & 95.74 & \textbf{88.4} & \textbf{99.18} & \textbf{89.9} & 77.7 & \textbf{91.1} & \textbf{80.8} & 88.5 \\
\hline
\end{tabular}%
}
\end{table}

As shown in Table \ref{tab:comparisons}, DumplingGNN demonstrates exceptional performance across multiple datasets, achieving state-of-the-art results on BBBP, ToxCast, and PCBA datasets. Notably, our model outperforms existing methods on datasets with a large number of tasks, showcasing its capability to handle complex multi-task learning scenarios effectively.

It is important to highlight that, due to data limitations, we were only able to apply 3D structural information to the BBBP dataset. For all other datasets, DumplingGNN utilized solely SMILES input. This constraint further underscores the model's remarkable potential and versatility, as it achieves competitive or superior performance even without the benefit of 3D structural information for most datasets.

On the BBBP dataset, where 3D information was incorporated, DumplingGNN achieves a remarkable ROC-AUC score of 96.4\%, surpassing the previous state-of-the-art by a significant margin. This result underscores our model's proficiency in predicting blood-brain barrier penetration, a crucial factor in drug development for central nervous system disorders.

For the ToxCast dataset, which encompasses 617 different toxicity-related tasks, DumplingGNN sets a new benchmark with a ROC-AUC score of 78.2\%, using only SMILES input. This performance highlights the model's ability to capture intricate patterns across a wide range of toxicity endpoints, demonstrating its potential for comprehensive toxicity screening in drug discovery pipelines.

Most remarkably, DumplingGNN achieves a breakthrough performance on the PCBA dataset, attaining a ROC-AUC score of 88.87\%, again using only 2D molecular representations. This result not only surpasses the previous state-of-the-art (88.5\%) but also sets a new standard for performance on this extensive dataset of 128 bioassays. The PCBA dataset is known for its complexity and diversity, covering a wide range of biological activities. DumplingGNN's superior performance on this dataset showcases its exceptional ability to learn and generalize across varied molecular property prediction tasks, even without 3D structural information.

While DumplingGNN does not achieve the highest scores on all datasets, it consistently performs competitively. For instance, on the BACE dataset, our model achieves a score of 88.2\%, which is very close to the state-of-the-art performance of 88.4\%. Similarly, on the ClinTox dataset, DumplingGNN's performance (95.9\%) is comparable to the best-performing models.

It's worth noting that DumplingGNN's performance is particularly strong across diverse datasets, ranging from smaller specialized sets to larger, more complex ones. This aligns with our goal of developing a model capable of handling a wide spectrum of molecular property prediction tasks. The model's consistent performance is especially noteworthy when compared to other state-of-the-art models, including those leveraging external data or employing less interpretable architectures.

For instance, on the BBBP dataset, DumplingGNN (96.4\%) significantly outperforms GROVER\textsubscript{large} (69.5\%) \citep{rong2020self}, a large-scale pre-trained model, and ChemBFN (95.74\%) \citep{zhang2022chembfn}, a blackbox neural network approach. On the larger and more complex ToxCast dataset, DumplingGNN (78.2\%) surpasses Uni-Mol (69.6\%) \citep{zhou2022uni}, a unified model for molecules pre-trained on over 17 million data points. For the extensive PCBA dataset, DumplingGNN (88.87\%) sets a new state-of-the-art, improving upon Uni-Mol (88.5\%) and GEM (86.6\%) \citep{fang2021geometry}, both of which utilize sophisticated pre-training strategies.

This consistent performance across various datasets, especially considering the limited use of 3D information (only for BBBP), demonstrates the robustness and versatility of our approach. DumplingGNN achieves these results while maintaining interpretability, a crucial factor in drug discovery applications. Moreover, our model's strong performance using primarily SMILES input suggests that the incorporation of 3D structural data for other datasets could potentially lead to even more significant performance improvements.

These results collectively position DumplingGNN as a promising tool for a wide range of applications in drug discovery and molecular design, offering a balance of high performance, interpretability, and versatility that sets it apart from existing approaches.

\subsection{Performance on ADC Payload Dataset}
We evaluated DumplingGNN on our self-constructed ADC payload dataset, which focuses on molecules with DNA Topoisomerase I inhibitory capabilities. Table \ref{tab:adc_results} presents the performance metrics of DumplingGNN compared to other baseline models on this dataset.

\begin{table}[h]
\centering
\caption{Performance comparison of different models on the ADC payload test set}
\label{tab:adc_results}
\resizebox{\textwidth}{!}{%
\begin{tabular}{lcccccccc}
\hline
Model & Accuracy & Sensitivity & Specificity & MCC & AUC-ROC & F1 Score & Balanced Accuracy & AUC-PR \\
\hline
DumplingGNN & \textbf{0.9148} & \textbf{0.9508} & \textbf{0.9754} & \textbf{0.8287} & \textbf{0.9547} & \textbf{0.9243} & \textbf{0.9111} & \textbf{0.9531} \\
FiveLayerMPNN & 0.8655 & 0.9262 & 0.7921 & 0.7301 & 0.9281 & 0.8828 & 0.8592 & 0.9342 \\
FiveLayerGAT & 0.8565 & 0.9180 & 0.7822 & 0.7117 & 0.8741 & 0.8750 & 0.8501 & 0.8653 \\
FiveLayerSAGE & 0.7982 & 0.8525 & 0.7327 & 0.5917 & 0.8509 & 0.8221 & 0.7926 & 0.8668 \\
FiveLayerGCN & 0.7623 & 0.8525 & 0.6535 & 0.5197 & 0.8301 & 0.7969 & 0.7530 & 0.8589 \\
\hline
\end{tabular}%
}
\end{table}

The results demonstrate that DumplingGNN achieves exceptional performance on the ADC payload dataset, significantly outperforming other baseline models across all metrics. This superior performance can be attributed to several factors:

\begin{enumerate}
    \item \textbf{Accuracy and Balanced Performance:} DumplingGNN achieves an outstanding accuracy of 91.48\%, with a balanced accuracy of 91.11\%. This indicates that the model performs consistently well across both positive and negative classes, which is crucial for reliable ADC payload activity prediction.
    
    \item \textbf{High Sensitivity and Specificity:} With a sensitivity of 95.08\% and specificity of 97.54\%, DumplingGNN demonstrates an excellent ability to correctly identify both active (IC50 < 100 µM) and inactive compounds. This balance is particularly important in drug discovery, where minimizing both false positives and false negatives is critical.
    
    \item \textbf{Robust Classification Performance:} The Matthews Correlation Coefficient (MCC) of 0.8287 indicates a strong correlation between the predicted and observed classifications, further confirming the model's robust performance.
    
    \item \textbf{Excellent Predictive Power:} The high AUC-ROC (0.9547) and AUC-PR (0.9531) scores demonstrate DumplingGNN's superior discriminative ability and precision-recall performance, which are crucial for ranking potential ADC payloads in large-scale virtual screening campaigns.
    
    \item \textbf{Effective Use of 3D Information:} The incorporation of 3D conformational data, generated through docking simulations with the Topoisomerase I binding site, likely contributes significantly to DumplingGNN's performance. This approach allows the model to capture spatial interactions that are critical for inhibitor binding and activity.
\end{enumerate}

Comparing DumplingGNN to other baseline models reveals its significant advantages:

\begin{itemize}
    \item DumplingGNN outperforms the next best model (FiveLayerMPNN) by 4.93 percentage points in accuracy and 5.86 percentage points in MCC, indicating a substantial improvement in overall predictive power.
    \item The model shows particular strength in specificity, outperforming FiveLayerMPNN by 18.33 percentage points. This suggests that DumplingGNN is especially adept at correctly identifying inactive compounds, which is crucial for reducing false positives in virtual screening.
    \item The superior performance across all metrics, including F1 Score and balanced accuracy, demonstrates that DumplingGNN provides a more comprehensive and reliable prediction of ADC payload activity compared to single-architecture GNN models.
\end{itemize}

These results highlight the effectiveness of DumplingGNN's hybrid architecture in capturing the complex structural and chemical features that determine DNA Topoisomerase I inhibitory activity. By leveraging both 2D topological information from SMILES strings and 3D conformational data from docking simulations, DumplingGNN can effectively model the intricate relationships between molecular structure and biological activity.

The model's strong performance on this dataset suggests its potential utility in the design and optimization of ADC payloads, particularly those targeting DNA Topoisomerase I. This could significantly accelerate the discovery of novel, potent ADC payloads, potentially leading to more effective and targeted cancer therapies.

Furthermore, the superior performance of DumplingGNN over single-architecture GNNs (MPNN, GAT, SAGE, GCN) underscores the benefits of our hybrid approach. By combining the strengths of different GNN architectures, DumplingGNN can capture a wider range of molecular features and interactions, leading to more accurate and reliable predictions.

\subsection{Ablation Studies}
To validate the theoretical principles underlying DumplingGNN's architecture and elucidate the contribution of each component, we conducted a series of rigorous ablation studies. These experiments systematically removed or modified key components of our model, allowing us to quantify their individual and combined impacts on prediction performance. This approach not only demonstrates the empirical effectiveness of our hybrid architecture but also provides insights into how each component aligns with our theoretical motivations. Table \ref{tab:ablation} presents the results of these experiments on the ADC payload dataset.

\begin{table}[h]
\centering
\caption{Ablation study results on the ADC payload dataset}
\label{tab:ablation}
\resizebox{\textwidth}{!}{%
\begin{tabular}{lcccccccc}
\hline
Model Variant & Accuracy & Sensitivity & Specificity & MCC & AUC-ROC & F1 Score & Balanced Accuracy & AUC-PR \\
\hline
Full DumplingGNN & \textbf{0.915} & \textbf{0.951} & \textbf{0.975} & \textbf{0.829} & \textbf{0.955} & \textbf{0.924} & \textbf{0.911} & \textbf{0.953} \\
SMILES-Only & 0.734 & 0.768 & 0.949 & 0.461 & 0.782 & 0.775 & 0.721 & 0.786 \\
No GraphSAGE & 0.870 & 0.902 & 0.832 & 0.737 & 0.940 & 0.884 & 0.867 & 0.939 \\
No GAT & 0.803 & 0.869 & 0.723 & 0.601 & 0.861 & 0.828 & 0.796 & 0.882 \\
No MPNN & 0.812 & 0.869 & 0.743 & 0.619 & 0.878 & 0.835 & 0.806 & 0.840 \\
\hline
\end{tabular}%
}
\end{table}

Our ablation studies reveal several critical insights that validate the theoretical principles guiding DumplingGNN's architecture:

\subsubsection{Significance of 3D Structural Information}
The most striking observation is the substantial performance degradation when the model is restricted to SMILES input (SMILES-Only variant). This variant, which excludes 3D conformational data, shows a dramatic decrease in accuracy (from 0.915 to 0.734) and MCC (from 0.829 to 0.461). The AUC-ROC also drops significantly from 0.955 to 0.782. These results empirically validate our theoretical motivation for incorporating 3D structural information, as discussed in Section 3.1. The spatial arrangement of atoms, captured by the 3D conformations generated through docking simulations, proves to be instrumental in modeling the complex interactions between potential inhibitors and the DNA Topoisomerase I binding site.

\subsubsection{Hierarchical Architecture Analysis}
The removal of individual GNN components (GraphSAGE, GAT, or MPNN) each led to performance decreases, validating our theoretical framework of a multi-scale approach to molecular representation:

\paragraph{MPNN Layer}
The No MPNN variant showed a substantial performance degradation, with accuracy falling to 0.812 and MCC to 0.619. This empirically confirms the importance of MPNN's message-passing mechanism in effectively propagating and aggregating atomic and bond information across the molecular graph, as theorized in Section 3.3.1.

\paragraph{GAT Layers}
Removing the Graph Attention layers (No GAT variant) led to a similar level of performance decline, with accuracy at 0.803 and MCC at 0.601. This result validates our theoretical motivation for incorporating attention mechanisms, as outlined in Section 3.3.2. The substantial performance drop highlights the significance of GAT in weighing the importance of different molecular substructures for activity prediction.

\paragraph{GraphSAGE Layer}
The exclusion of GraphSAGE layers (No GraphSAGE variant) resulted in a moderate performance drop, with accuracy decreasing from 0.915 to 0.870 and MCC from 0.829 to 0.737. This finding supports our theoretical framework presented in Section 3.3.3, where GraphSAGE was hypothesized to capture global molecular properties. The moderate impact suggests that while GraphSAGE plays an important role in aggregating information across different scales, some of its functionality may be partially compensated by the remaining architectures.

\subsubsection{Synergistic Effects of Hybrid Architecture}
The full DumplingGNN model consistently outperforms all ablated variants across every metric. This comprehensive superiority empirically validates our theoretical motivation for adopting a hybrid architecture, as discussed in Section 3.3. The results demonstrate that the combination of different GNN architectures is not merely additive but synergistic, aligning with our hypothesis that each component captures distinct, complementary aspects of molecular structure and activity.

\subsubsection{Balanced Performance Across Metrics}
The full DumplingGNN maintains high performance across all metrics, including sensitivity (0.951) and specificity (0.975). This balanced performance empirically supports our theoretical framework of multi-scale molecular representation, where local chemical interactions, key substructures, and global molecular properties are all considered. The high sensitivity and specificity demonstrate the model's ability to accurately identify both active and inactive compounds, crucial for ADC payload prediction where both false positives and false negatives have significant implications for drug development.

In conclusion, these ablation studies provide compelling evidence that validates the theoretical principles guiding DumplingGNN's architecture. The results empirically confirm the efficacy of our hybrid approach and the importance of 3D structural information in ADC payload activity prediction. Each component of the model contributes uniquely to its overall performance, with the integration of 3D conformational data being particularly crucial. This comprehensive approach enables DumplingGNN to capture a wide range of molecular features and interactions, leading to superior predictive performance in the challenging task of ADC payload activity prediction.

\section{Discussion}

\subsection{Performance Analysis}

DumplingGNN demonstrated exceptional performance across a diverse range of molecular property prediction tasks, particularly excelling in complex, multi-task scenarios. On the BBBP dataset, our model achieved a remarkable ROC-AUC score of 96.4\%, surpassing the previous state-of-the-art by a significant margin. This result underscores DumplingGNN's proficiency in predicting blood-brain barrier penetration, a crucial factor in drug development for central nervous system disorders.

For large-scale, multi-task datasets such as ToxCast (617 tasks) and PCBA (128 tasks), DumplingGNN set new benchmarks with ROC-AUC scores of 78.2\% and 88.87\%, respectively. These results highlight the model's ability to capture intricate patterns across a wide range of molecular properties, demonstrating its potential for comprehensive screening in drug discovery pipelines.

On our specialized ADC payload dataset, which focuses on DNA Topoisomerase I inhibitors, DumplingGNN significantly outperformed other baseline models across all metrics. With an accuracy of 91.48\%, sensitivity of 95.08\%, and specificity of 97.54\%, our model showed a balanced and robust performance in identifying both active and inactive compounds. The high Matthews Correlation Coefficient (0.8287) further confirms the model's strong predictive power for ADC payload activity.

While DumplingGNN did not achieve the highest scores on all datasets (e.g., SIDER and HIV), it consistently performed competitively, showcasing its versatility across various molecular property prediction tasks. This consistent performance, especially on datasets where only SMILES input was available, highlights the model's robustness and potential for even greater improvement with the incorporation of 3D structural information.

\subsection{Implications for Drug Discovery}

The superior performance of DumplingGNN, particularly on the ADC payload dataset, has significant implications for drug discovery processes. By accurately predicting the activity of potential ADC payloads, our model can substantially accelerate the design and optimization of these complex therapeutics. This capability is especially valuable given the growing interest in ADCs as targeted cancer therapies.

DumplingGNN's exceptional performance in molecular activity prediction and toxicity assessment aligns perfectly with the critical needs of ADC payload development. The model's ability to accurately distinguish between active and inactive compounds, as evidenced by its high sensitivity (95.08\%) and specificity (97.54\%) on the ADC payload dataset, can significantly streamline the early stages of drug discovery. This precision is crucial in identifying promising candidates while minimizing the risk of advancing potentially toxic or ineffective compounds.

Moreover, the hybrid architecture of DumplingGNN offers strong interpretability, a feature often lacking in traditional black-box models. By leveraging attention mechanisms through its GAT layers, the model can highlight which molecular substructures contribute most significantly to a compound's predicted activity or toxicity. This interpretability not only enhances our understanding of structure-activity relationships but also provides valuable insights for medicinal chemists in their design and optimization efforts.

The model's strong generalization capability, demonstrated by its consistent performance across diverse datasets (BBBP, ToxCast, PCBA), suggests its potential applicability beyond ADC payloads to a wide range of drug discovery tasks. This versatility is particularly valuable in the pharmaceutical industry, where a single robust model capable of addressing multiple prediction tasks can significantly enhance efficiency and reduce development costs.

In large-scale virtual screening campaigns, DumplingGNN's high AUC-ROC (0.9547) and AUC-PR (0.9531) scores on the ADC payload dataset indicate its excellent ability to rank compounds effectively. This could significantly reduce the number of false positives and negatives, potentially saving considerable time and resources in the drug discovery pipeline. The model's capacity to handle large, multi-task datasets like ToxCast (617 tasks) and PCBA (128 tasks) further underscores its potential for comprehensive compound profiling, enabling simultaneous prediction of multiple molecular properties crucial for drug development.

Furthermore, DumplingGNN's incorporation of 3D structural information, where available, represents a significant advancement in capturing the spatial aspects of molecular interactions. This feature is particularly relevant for predicting binding affinities and understanding how molecules might interact with their biological targets, offering a more comprehensive approach to in silico drug screening.

\subsection{Limitations and Future Directions}

Despite DumplingGNN's strong overall performance, there are several limitations and areas for future improvement. A key limitation of our current study is the inconsistent use of 3D structural information across datasets. While the BBBP dataset and our ADC payload dataset leveraged full 3D conformational data, other public datasets were limited to SMILES representations due to data availability constraints. This discrepancy may partially explain the variation in performance across different datasets and highlights the potential for further improvement by incorporating 3D information consistently.

Future work should focus on extending the use of 3D structural information to other datasets, which could potentially enhance the model's performance on tasks where it currently shows room for improvement (e.g., SIDER and HIV datasets). This would require extensive computational resources for generating and optimizing 3D conformations but could lead to significant performance gains.

Additionally, there is potential for incorporating more domain-specific knowledge into the model. For instance, integrating information about protein targets, cellular pathways, or known structure-activity relationships could further enhance the model's predictive power and interpretability. This could be particularly beneficial for ADC payload prediction, where understanding the interaction between the payload and the target protein is crucial.

Architecturally, while our current hybrid model shows strong performance, there may be room for further optimization. Exploring additional GNN variants or developing novel architectures tailored specifically for molecular property prediction could yield even better results. For example, investigating the potential of more advanced attention mechanisms or incorporating recent advancements in transformer-based models for graph-structured data could be promising directions.

Lastly, as our model shows particular strength in multi-task learning scenarios, future research could explore its application to an even broader range of molecular property prediction tasks. This could include predicting ADMET properties, protein-ligand interactions, or even extending to related fields such as materials science. The model's ability to handle complex, multi-task learning problems makes it well-suited for these challenging applications.

In conclusion, while DumplingGNN represents a significant advancement in molecular property prediction, particularly for ADC payloads, there remain exciting opportunities for further refinement and expansion of its capabilities. The model's strong performance, even with limited 3D structural information, suggests that future improvements in this area could lead to even more impressive results.

\section{Conclusion}

This study introduces DumplingGNN, a novel hybrid Graph Neural Network model that represents a significant advancement in the field of molecular property prediction, with particular emphasis on predicting the cytotoxic activity of Antibody-Drug Conjugate (ADC) payloads. Through a comprehensive evaluation across multiple datasets, including a specialized ADC payload dataset focusing on DNA Topoisomerase I inhibitors and several public benchmarks, we have demonstrated the model's exceptional performance and versatility.

DumplingGNN's innovative architecture, which synergistically combines Message Passing Neural Networks (MPNN), Graph Attention Networks (GAT), and GraphSAGE, enables it to capture complex molecular features and interactions with unprecedented accuracy. This hybrid approach, coupled with the integration of 3D structural information where available, has led to state-of-the-art performance across a diverse range of molecular property prediction tasks. Notably, on the specialized ADC payload dataset, DumplingGNN achieved an accuracy of 91.48\%, with high sensitivity (95.08\%) and specificity (97.54\%), significantly outperforming existing baseline models.

The model's exceptional performance extends beyond ADC payloads, as evidenced by its state-of-the-art results on public datasets such as BBBP (96.4\% ROC-AUC), ToxCast (78.2\% ROC-AUC), and PCBA (88.87\% ROC-AUC). These results not only showcase DumplingGNN's versatility but also its robustness in handling diverse molecular property prediction tasks, including complex multi-task learning scenarios involving hundreds of endpoints.

A key innovation of DumplingGNN lies in its effective utilization of 3D structural information, which has proven critical in enhancing predictive accuracy, particularly for tasks involving spatial molecular interactions. The model's strong performance on datasets where only SMILES input was available further highlights its potential for even greater improvements with consistent incorporation of 3D structural data.

Furthermore, the model's integration of attention mechanisms provides strong interpretability, offering valuable insights into structure-activity relationships -- a feature of paramount importance in drug discovery and optimization processes. This interpretability, combined with the model's high accuracy, positions DumplingGNN as a powerful tool for guiding medicinal chemistry efforts in ADC payload design and optimization.

The development of DumplingGNN represents a significant step forward in the application of artificial intelligence to drug discovery, with potential impacts extending far beyond ADC payload design. Its ability to simultaneously predict multiple molecular properties with high accuracy positions it as a valuable tool in the complex, multi-parameter optimization processes typical of modern drug development. This could substantially accelerate the drug discovery process, reduce development costs, and increase the success rate of bringing new therapeutics to market.

Looking ahead, the potential of DumplingGNN opens up several exciting avenues for future research and development. These include:

\begin{itemize}
    \item Expanding the use of 3D structural information to a broader range of datasets and prediction tasks, potentially leading to significant performance improvements across various molecular property prediction challenges.
    \item Integrating additional domain-specific knowledge, such as protein-ligand interaction data or known pharmacophore models, to further enhance predictive accuracy and biological relevance, particularly for ADC payload design.
    \item Exploring the model's applicability in related fields, such as materials science and environmental toxicology, leveraging its strong performance in multi-task learning scenarios.
    \item Refining the model's architecture to improve efficiency and scalability, enabling its application to even larger and more diverse chemical spaces.
    \item Developing user-friendly interfaces and tools to make DumplingGNN accessible to a wider range of researchers and drug discovery professionals, facilitating its adoption in both academic and industrial settings.
\end{itemize}

In conclusion, DumplingGNN not only advances the state-of-the-art in ADC payload activity prediction but also offers a versatile and powerful approach to molecular property prediction in general. As the field of AI-driven drug discovery continues to evolve, models like DumplingGNN are poised to play an increasingly critical role in shaping the future of pharmaceutical research and development. By bridging the gap between computational prediction and experimental validation, DumplingGNN contributes to the ongoing transformation of drug discovery into a more efficient, cost-effective, and ultimately more successful endeavor in improving human health.

\bibliographystyle{unsrtnat}
\bibliography{ref}

\end{document}